\documentstyle[12pt]{article}
\begin{document}
\parindent 1.3cm
\thispagestyle{empty}   
\vspace*{-3cm}
\noindent
\def\sd{\strut\displaystyle}
\renewcommand{\thefootnote}{\fnsymbol{footnote}}

\begin{obeylines}
\begin{flushright}
UB-ECM-PF 94/13
\end{flushright}
\end{obeylines}

\vspace{2cm}

\begin{center}
\begin{bf}
\noindent
 Enhancing the top signal at Tevatron using Neural Nets \end{bf}
 \footnote{This research is partly supported by EU under contract number
 CHRX-CT92-0004.}

  \vspace{1.5cm}
Ll. Ametller$^a$, Ll. Garrido$^{b,c}$ and P. Talavera$^a$
\vspace{0.5cm}\\

$^a$Departament F{\'\i}sica i Enginyeria Nuclear\\
Universitat Polit\`ecnica de Catalunya, E-08034 Barcelona, Spain\\

\vspace{0.3cm}
$^b$Departament Estructura i Constituents Mat\`eria\\
Universitat de Barcelona, E-08028 Barcelona, Spain\\

\vspace{0.3cm}
$^c$Institut de F{\`\i}sica d'Altes Energies, Universitat Aut\`onoma de
    Barcelona\\
    E-08193 Bellaterra (Barcelona), Spain\\

\vspace{4cm}

{\bf ABSTRACT}
\end{center}

We show that Neural Nets can be useful for top analysis at Tevatron.
The main features of $t\bar t$ and background events on a mixed sample
are projected in a single output, which controls the efficiency and purity
of the $t\bar t$ signal.

\newpage
The announced discovery of the top quark by CDF \cite{CDF} at Tevatron has
originated a
big excitation in the scientific community. Although the statistics is too
limited \footnote{CDF has reported on 12 events, with 6 events for the
estimated background, with a $0.26\%$ probability of observing
background fluctuation.  D0 instead has not a
clear signal of the top quark \cite{D0}.}
to establish the existence of the top quark, it is however natural to
interpret the excess of events as $t\bar t$. The experimental situation
will certainly improve in next months and top will hopefully be confirmed.
 From the theoretical point of view, the consistency of the Standard Model
demands top to be the partner of the
bottom quark, ensuring the absence of flavor changing neutral currents
\cite{CLEO}.
The CDF value of the top mass, $m_t=174\pm10^{+13}_{-12}$ GeV
\cite{CDF}, is consistent with recent theoretical studies on
radiative corrections combined with precision measurements of the Z boson
mass and the strong coupling constant at LEP leading to
$m_t=165^{+13+18}_{-14-19}$ GeV \cite{Miquel}.

The dominant top production mechanism at Tevatron is $q \bar q \to t \bar
t$, followed by  $g g \to t \bar t$.
Once produced, the top decays into $b W$, with the subsequent $W\to
l\nu,q\bar q'$ decay, in the detector. There are therefore three possible
final states for the $t \bar t$ signal which , on increasing branching
ratios, are:

\begin{enumerate}
       \item{}   Two charged leptons, missing energy and two jets
       \item{}   One charged lepton, missing energy and four jets
       \item{}   Six jets.
\end{enumerate}

They need different strategies for top searches and different backgrounds
have to be considered respectively.
The first channel suffers from a small branching ratio and the presence of two
undetected neutrinos that makes top reconstruction unfeasible. It has been
analyzed in terms of the correlations among the charged leptons \cite{BOP}
and, recently, it has been suggested to be separable from its possible
backgrounds \cite{HP}.
The most investigated channel so far is the one containing one charged
lepton \cite{Single}. It has a sizeable branching ratio with a moderate
background. Still the neutrino escapes detection and hence the event can not
be completely reconstructed.
The third channel, six final jets, is the most likely and allows full
top reconstruction but at expenses of a huge QCD background. Recently, it
has been pointed out that tagging of a b-quark can help to obtain
acceptable signal to background ratios for $m_t<180$ GeV \cite{Giele}.

All mentioned channels need some specific experimental cuts for detecting
jets and/or hard leptons as well as for their isolation. This, together
with detector performance, implies a sensible reduction on the number of
possible $t \bar t$ candidates, and demands a good efficiency for
discerning real  from fake $t \bar t$ background events. We propose to use
Neural Nets (NNs) for the analysis of experimental data trying to maximize
the signal to background ratio without significant loses in statistics,
in particular to top analysis at Tevatron.
NNs are by now well known for its ability in classifying
among different distributions and are being used for this purpose in
several high energy applications \cite{Vicens}. Some examples are Higgs
search at LHC \cite{ChSt}, b and $\tau$ analysis \cite{ArgusTau}, quark
and gluon jets analysis \cite{Jets}, determination of Z to heavy quarks
branching ratios \cite{Z}, or bottom jet recognition \cite{Odorico}.
It has been shown that NNs give, after
proper training, the probability that a given event belongs to some class
\cite{Lluis} providing therefore a useful tool for classifying decisions.
In
fact, we are not interested in a deep and exhaustive analysis but rather in
the possibilities that a NN can offer us for enlarging the signal to
background ratio. For we restrict ourselves  at
the parton level, without considering hadronization, detector acceptance,
resolution effects, efficiencies, etc. in order to illustrate the
potential effects of the NN in front of the classical analysis in terms of
cuts on a given set of variables.

We focused our analysis to the one charged lepton channel
\begin{equation}
\label{signal}
 p \bar p\to t \bar t \to l \nu j j j j,
\end{equation}
with $l= e^{\pm}, \mu^{\pm}$, using the exact tree level amplitudes with
spin correlations \cite{KS}. The main background to this process is
\cite{Wjets}
\begin{equation}
\label{mainback}
 p \bar p \to W j j j j \to l \nu j j j j
\end{equation}
together with
\begin{equation}
\label{smallback}
p \bar p \to W W (W Z) j j \to l \nu j j j j
\end{equation}
which is an order of magnitude smaller \cite{WW}. We have only considered
the first mechanism and have used VECBOS\footnote{We thank W. Giele for
making the VECBOS code available to us.} \cite{Vecbos} for its evaluation.

We have taken  $m_t=174$ GeV and have normalized the
total $t \bar t $ cross section at Tevatron to $5.1$~pb, value that takes
into account $O(\alpha_s^ 3)$ corrections and resummation of leading soft
gluon corrections to all orders in perturbation theory \cite{Laenen}.
CDF measures a $t\bar t$ cross section of $13.9^{+6.1}_{-4.8}$~pb \cite{CDF}
which is a factor around $2.5$ bigger than the theoretical value we have used.
Notice that using the CDF value, the signal to background ratio
would increase by the same factor.
We have used the HMRS set~1 structure functions \cite{HMRS} at the scale
$Q=m_t$ ($Q= <p_t>$) for the top signal (background). We generated events
satisfying reasonable acceptance cuts for the jets, charged lepton and
missing transverse momentum,
\begin{equation}
\label{ptcut}
p_T^j, p_T^l, p\llap /_T > 20   \  \hbox{\rm GeV},
\end{equation}
and the jets and lepton pseudorapidities
\begin{equation}
\label{etacut}
|\eta^j|, |\eta^l| < 2,
\end{equation}
and requiring jet and lepton isolation,
\begin{equation}
\label{rcut}
\Delta R_{jl}, \Delta R_{j j} > 0.7,
\end{equation}
where $\Delta R=\sqrt{(\Delta \eta)^2+(\Delta \phi)^2}$ is the distance in
the lego plot. These cuts are intended to simulate the experimental cuts
needed to detect jets and hard leptons inside the detector and to select
good candidates for top production
(from now on these cuts will be referred to as acceptance cuts).
 The cross section after the  acceptance
cuts is $0.35$ pb ($1.2$ pb) for $t \bar t$ signal (background) in good
agreement with Ref.\cite{BOP2}. We generated $4000$ $t\bar t$ and $4000$
background events.
The total number of events is essentially limited by the time needed to
generate a statistically significant sample for the background. (More
efficient generation techniques have been recently proposed \cite{Giele}
wich could hopefully circumvent this problem).

Notice that the acceptance  cuts have to be supplemented either with
additional cuts or any other criteria, as a NN for instance, on some
kinematical variables in order to assign a single event as signal or
background, leading to a reduction of the $t \bar t$ and background
event samples. (b tagging, for example, reduces the signal by a factor of
order $0.3$.\cite{Proto})

We have considered six kinematical variables in our analysis,

\begin{obeylines}
\begin{itemize}
\item{} $i$) $p_T^{W_l}$, the transverse momentum of the leptonically decaying
$W$.
\item{} $ii$) $E_T$, the total transverse energy.
\item{} $iii$) $m_{W_{jj}}$, the invariant mass of the hadronically decaying
$W$.
\item{} $iv$) $m_t$, the reconstructed top mass.
\item{} $v$) $S$, sphericity.
\item{} $vi$) $A$, aplanarity.
\end{itemize}
\end{obeylines}

Variables $i$ and $ii$ are completely defined when assigning the missing
transverse momentum to the undetected neutrino. The third variable
requires pairing of two jets with invariant mass close to the $W$ mass.
Variables $iv$, $v$ and $vi$ need the knowledge of the longitudinal
momentum of the neutrino, which is not measured. It can however be
inferred assuming that the $l\nu$ pair comes from an on--shell $W$. This
leads to a two--fold ambiguity which can be resolved to some extend by
requiring $t \bar t$ reconstruction in the lines suggested by Ref.\cite{BOP2}
to which we refer for details. The sphericity and aplanarity, computed for
the lepton plus neutrino plus 4-jet momenta, take into
account the topology of the events expecting larger values from the signal
than the background distributions.

The usual strategy for classifying signal or background type events is by
applying different cuts on the kinematical variables considered, the six
above mentioned in our case.
These cuts are usualy given by simple expressions (for instance:
var1 $>$ cut1 and var2 $<$ cut2), so that, the different regions are
separated by hyperplanes in the variable space (from now on
these cuts will be referred to as kinematical cuts).
Denoting by $T$ ($B$) the number of top
signal (background) events passing our selection criteria, and $T_t$ the
total number of $t \bar t$ events selected after the acceptance cuts,
Eqs.(4--6), one would like to
find
the best combination of cuts on the kinematical variables
such to maximize the efficiency $\eta\equiv
T/T_t$ or the purity $P\equiv T/(T+B)$ or both simultaneously.
In the latest case, a method could be to
maximize the statistical significance of the filtered subsample, $S_s\equiv
T/\sqrt{B}$, criterium that can be used to enhance a new signal from
its expected
background.
In any case, this gives rise to subtle fine tuning on the
cuts to reach the maximization that can become a hard issue
for larger number of kinematical variables considered.

We are interested in the separation of signal and background  using
a layered feedforward NN which, as we will show,
avoids fine tuning in a multi variable
space. A feedforward NN consists of several layer of units called neurons.
Between the layer we can distinguish
one input layer where the information comes in, one or several
hidden layers where the information is processed, and one
output layer which yields the output of the NN.

The input of neuron $i$ in layer $l$ is given by,
\begin{eqnarray}
\label{neusum}
I^{l}_{i} & = & \sum _{j} w_{ij}^{l} S_{j}^{l-1}+ B_{i}^{l} \;\;\; l=2,3,... \\
I^{1}_{i} & = & in^{(e)}_{i} \;\;\; ,
\end{eqnarray}
where $in^{(e)}_{i}$ is the set of kinematical variables for event $e$,
the sum is extended over the neurons of the preceding layer
$(l-1)$, $S_{j}^{l-1}$ is the state of the neuron $j$,
$w_{ij}^{l}$ is the connection weight between the neuron
$j$ and the neuron $i$, and  $B_{i}^{l}$ is a bias input to neuron $i$.
The state of a neuron is a function of its input
$S_{j}^{l}=F(I^{l}_{j})$, where $F$ is the
neuron response function.
In this paper we  take
$F(I^{l}_{j})=1/(1+exp(-I^{l}_{j}))$, the so-called ``sigmoid function'',
 which is similar to the response curve of the biological
neuron and offers more sensitive modelling
of real data than a linear function.

The parallel behaviour of NNs has the
capacity of learning over a set of given examples.
A very popular learning algorithm is the error
backpropagation (BP) \cite{BP}. The main objective
of the BP is to minimize an error function, also
called energy
\begin{equation}
\label{energy}
E=E(in^{(e)}_i,out^{(e)},w_{kl},B_n)=
\frac{1}{2}\sum_{e}(o^{(e)}-out^{(e)})^2 \ ,
\end{equation}
by adjusting the $w_{kl}$ and $B_n$ parameters and  being
$o^{(e)}$
the state of the output neuron, $out^{(e)}$ its desired state,
and $e$ runs over the event sample.  Taking the desired output
as 1 for each signal event and 0 for each background event, the
output of the net, after training, gives the conditional probability
that given the observed quantities for a single event, this event is a signal
\cite{Lluis}, provided that the ratio of signal to background in the learning
sample corresponds to the real one.

We have used a 3 layer NN with 6 input neurons that are activated
with the kinematical variables mentioned in the previous section
(normalized to 1 for convenience), a
hidden layer with 6 neurons, and a unique output neuron which desired
output is 1 for the signal and 0 for the background. We have found that
using 6 neurons in the hidden layer optimizes the minimum energy.

For the training step we have used 2000 top events and 2000
background events which do not correspond to the expected cross sections
ratio. However since we are not interested in the conditional probability
mentioned above but to study the efficiency and purity as a function of
the cut on the output activation of the NN, this fact will not produce
any trouble and the learning results more efficient.
As a test sample, we have taken  570 (2000) top (background)
events statistically independent from the training ones. The top/background
ratio of the test sample is chosen equal  to the obtained from the
expected cross sections.
All results
presented have been obtained from the test sample.

Figure 1 shows the distribution of signal and background events as a function
of the NN output activation for the test sample.
We see two  peaks close to 1  and 0 corresponding mainly to the signal
and background respectively. It is clear from this plot that
cutting on the output of the net we can have samples richer on signal
or in background as desired.

Solid (dashed) line in Figure 2 shows the efficiency (purity) as a function
of the net output cut. It is clear that we have to choose an
output cut close to 1 if we want high purity or a cut close to 0 for high
efficiency.
The highest output cut to improve the purity, given a fixed luminosity, would
be the one leading to still enough signal events (as minimum $5$). This cut
will be very close to 1, due to the fact that
the efficiency is larger than $0.9$ for any value of the output cut
except for values very close to $1$,

Figure 3 shows the efficiency versus the purity
(solid line)  when varing the NN output cut from 0 to 0.99998.
The points correspond to some hypercubic
cuts applied over the six input variables, and have to be considered as
the traditional procedure (each point represents a given combination of
cuts bigger than certain values, or masses located around a certain central
value, for instance $p_t > p_t^{min}$, $S>S^{min}$, $m_W - \delta <
m_{W_{jj}} < m_W+\delta$,...), chosen favoring the signal in front of the
background.
We find that the NN performance, working only with one variable, the output
of the net, is better than the traditional analysis for
any combination of purity and efficiency, showing the great improvement of
the method.  A complex problem on many variables has been reduced to
the study of only one variable, the NN output, which even improves the
analysis.

When the important fact is to reveal the existence of the signal the relevant
quantity should be the statistical significance.
Values of $S_s > 5$  are
commonly accepted as a proof of the existence of a clear signal.
Figure 4 shows the relation of the statistical significance versus the
efficiency and the purity for $T_t=1$ signal events
 (changing
the number of signal events, $T_t$, the surface in Fig. 4 does not modify its
shape and only
rescales its height which is proportional to $\sqrt{T_t}$).
Figure 5 shows the statistical significance as a function of the net output
for $7$ signal events before kinematical cuts
(corresponding to an
integrated luminosity of $20$~pb$^{-1}$).
We see that the statistical significance increases as the output cut increases.
As in the case for improving purity, the highest output cut, given a fixed
luminosity, would
be the one leading to still enough signal events (as minimum $5$), and is
very close to $1$.

One of the problems that is faced in $p\bar p$ collisions is the estimation of
the background. A factor $2$ on the background could destroy any
evidence of the signal. In Figure 6 we have the allowed region
of the output cut versus the factor $f$ of the background ( $f=2$ means
 that the background is two times bigger as we have computed) where, for the
 luminosity of $20$~pb$^{-1}$, we still can obtain a 5 sigma effect with
at least $5$ signal events. Given
a fixed factor $f$ the largest and smallest  values of the output cut
correspond to the  highest purity and highest efficiency respectively.
Notice that  output cuts very close to 1 are not included in the allowed
region, although this is not visible in the plot.


Our results indicate that NNs are suitable for top analysis at Tevatron.
Although we focused our study in a particular channel and
worked at the parton level, we expect similar behaviour for the other
channels with the corresponding backgrounds and when performing
more realistic analysis including hadronization and detector simulation.
We do not claim to have used neither the best kinematical variables for
our analysis, nor to find the best NN topology. Our aim was only to study
the potential use of NNs as a cross check to the traditional analysis in
terms of cuts on a multidimensional variable space. More elaborated
studies are postponed for a forthcoming publication.

In conclusion, we have shown that a NN trained with a mixed sample
of $t\bar t$ and background events learns the main features of the
different samples in a multivariable input space and projects them in a
single output. This output turns out to be very useful for discrimination
between signal and background events.

\vskip0.5truecm

{\it Acknowledgements.}

We thank F. Aguila and G. Stimpfl for discussions.

\newpage

{\bf Figure Captions}

\begin{itemize}
\item Fig.1  Distribution of the signal (dashed)
 and background (solid) events as a function
of the NN output activation for the test sample
consisting on 570 (2000)
top (background) events. Values close to 1 (0)  correspond mainly  to  top
(background) events.

\item Fig.2  Efficiency (solid line) and purity (dashed line) as a function of
the NN output cut.

\item Fig.3  Efficiency versus purity for the test sample. The solid line
shows the NN result whereas the points correspond to several sets of linear
 cuts (see text)
applied to the six input variables.

\item Fig.4  Statistical significance as a function of the efficiency and
purity normalized to $T_t=1$ signal events. It scales as $\sqrt{T_t}$.

\item Fig.5  Statistical significance as a function of the NN output cut for an
integrated luminosity of $20$~pb$^{-1}$.

\item Fig.6  Allowed region (shaded area)
of the output cut versus the factor $f$ of the background ($f=2$ means
 that the background is two times bigger as we have estimated) where, for the
 luminosity of $20$~pb$^{-1}$, we still can obtain a 5 sigma effect with
at least $5$ signal events.

\end{itemize}

\end{document}